\magnification 1200 
\centerline{\bf  Market Ecology, Pareto Wealth Distribution and 
Leptokurtic Returns in }
\centerline{\bf  Microscopic Simulation of the LLS Stock 
Market Model} 

  \bigskip 
\centerline{\bf Sorin Solomon $^1$ and Moshe Levy $^2$} 

  \bigskip 

\centerline{\bf $^1$ Racah Institute of Physics, Givat Ram 91104, }
\centerline{\bf $^2$ School of Business Administration, 
Mount Scopus 91905, }
\centerline{\bf Hebrew University of Jerusalem }

  \bigskip
{\centerline {\bf Abstract:}} 

The LLS
 stock market model is a model of heterogeneous
quasi-rational investors operating in a complex environment
about which they have incomplete information. We review the main
features of this model and several of its extensions. We study the effects
of investor
heterogeneity and show that predation, competition, or symbiosis may occur
between different investor populations. The dynamics of the LLS model
lead to the empirically observed Pareto wealth distribution.
Many properties observed in actual markets appear as natural consequences
of the LLS dynamics: truncated Levy distribution of short-term returns,
excess volatility, a return autocorrelation "U-shape" pattern, and a
positive correlation
between volume and absolute returns.

   {\bf 1. The LLS Model } 

LLS is a microscopic representation model of the stock market. 
Its details and some generalizations of it
can be found in [2]. In the present account we introduce the 
basic LLS ideas and the model main results. 
We consider below a market with only two investment options: 
a bond and a stock (see [3] for an extension to a 
multiple stocks case). 
The model involves a large number of virtual investors characterized 
each by a current wealth, portfolio structure,
probability expectations and risk taking preferences. These 
personal characteristics come into play in each investor's decision 
making process as schematically seen in Fig. 1.

The bond is assumed  to be a risk-less asset yielding a return at 
the end of each time period. The bond 
is exogenous and investors can buy from it as much as they wish 
at a given rate. 

The stock is a risky asset with overall returns rate $H(t)$ 
composed of two elements: 
\medskip 
(i). Capital gain (loss): 
If an investor holds a stock, any rise (fall) in the 
market price of the stock contributes to 
an increase (decrease) in the investors' wealth. 
\medskip 
(ii). Dividends: The company earns income and distributes dividends. 
\medskip 

Each investor $i$ is confronted with a decision where the outcome is
uncertain: which is the optimal fraction $X(i)$  
of his/her wealth to invest in stock?
 According to the standard theory of investment each 
investor is characterized by a utility function (of its wealth) 
$U(W )$ that reflects his/her personal risk 
taking preference (here we take for simplicity $U(W)= \ln W$ 
see [1]
for a prospect theory extension). The optimal $X(i)$ is the one that 
maximizes the expected value of his/her $U(W)$ 
(we take into account all the unknown factors influencing 
decision-making (such as liquidity constraints or deviations from 
rationality) by adding a small random variable (or "noise") to 
the optimal proportion $X(i)$). 
The expected value of $U(W)$ depends of course 
on the expected probabilities for the various values of $H$ to be 
realized in the future. In LLS the investors
expectations for the future $H$'s are based on extrapolating the 
past values. More precisely 
each investor recalls the last k returns on the stock and expects 
that each of them may take place again 
with equal probability. The extrapolation range k differs between 
various investors and it will be in the 
sequel of this paper the main parameter inducing market inhomogeneity.

At fixed time intervals each investor revises the composition of its 
portfolio and decides for a new market order. 
The aggregate of these orders determines the new stock price 
by the market clearance condition as explained below.
Once each investor decides on the proportion of his/her wealth 
$X(i)$ 
that (s)he wishes to hold in stocks, one can derive the number of 
stocks 
$N(i,p_h)$ it wishes to hold corresponding to each hypothetical stock 
price $p_h$. 
Since the total number of shares in the market $N$,   
is fixed there is a particular value of the price $p$ for 
which the sum of the $N(i,p)$ equals $N$. 
This value $p$ is the new market equilibrium price. 
Upon updating accordingly the traders' portfolios, 
wealth and list of last $k$ returns, one is ready for the
next market iteration. This process is repeated for each time step,
and the market prices are recorded throughout the run.
\medskip 
{\it Figure 1: The Flow Chart of the LLS market framework}

  \bigskip 
   {\bf 2. Crashes, Booms and Cycles } 

The LLS model provides already at the level of a quite 
homogenous 
tradersÒ population a convincing 
description of the emergence of cycles of booms and crashes 
in the stock markets. In a market with one 
species of investors all having a homogenous memory (extrapolation) 
range spanning the last k returns of the stock, 
the stock price alternates regularly between two very different price
levels. The explanation for this 
behavior is as follows: 

Assume that the rate of return $H(t)$
on the stock at a time $t$ is higher than the oldest remembered return 
($H(t-k)$). 
The addition of $H(t)$ and the elimination of $H(t-k)$ creates then 
a new distribution of past returns that is better 
than the previous one. 
Since the LLS extrapolating investors use the past $k$ 
returns to estimate the distribution of the next period's return, 
they will be lead to be more optimistic and increase their 
investments in the stock. 
This, in turn, will cause the stock price to rise, which will generate 
an even higher return. 

This positive feedback loop stops only when investors reach the 
maximum investment proportion 
(i.e. $X(i)=100 \%$: we do not allow borrowing or short selling), 
and can no longer increase their investment proportion in the stock. 
The dividend contribution 
to the returns is small compared with this high  price at this stage. 
In the absence of noise the returns on 
the stock at this plateau converge to a constant growth rate which 
is just slightly higher than the riskless 
interest rate (see [4]). 
In other words, in the absence of noise the price remains almost 
constant, growing only 
because of the interest paid on the bond (more money entering the 
system and being invested in the stock). 

When there is some noise in the system the price fluctuates a little 
around the asymptotic high level, 
because of the small random fluctuations in the investment proportions. 
These fluctuations generate some negative 
returns (on a downward fluctuation) and some high returns 
(when the price goes back up). One might 
suspect that a large downward fluctuation might trigger a reverse 
positive 
feedback effect, trader expectations will lower, 
investment proportions will decrease, the price will drop, 
generating further negative returns and so on:
a crash.  This can happen during the "plateau" period but only after 
the previous sharp price boom which generated  an extremely 
high return, is forgotten. And, indeed, this is exactly what happens. 
Since it takes $k$ steps to forget the 
boom, the high price plateaus are a bit longer than the extrapolation 
span ($k$ days to forget the boom 
$+O(1)$  more days until a large enough negative fluctuation occurs). 

The crash generates a disastrous return and, until it is forgotten, 
investment proportions and hence the 
price remains very low. When the price is low, the dividend becomes 
significant and the returns on the 
stock are relatively high (compared with the bond). 
Once the crash has been forgotten, all the returns that are remembered
 are 
therefore high, and the price jumps back up. Thus, the low price 
plateaus are $k$ steps long. This 
completes one cycle, which is repeated throughout the run. 
This (quasi-)periodicity is best viewed in the 
Fourier transform of the price time evolution (Fig. 2)  
as a series of narrow peaks around the frequency 
$2k+O(1)$ and its harmonics 
(note however that the dynamics is not perfectly periodic 
and therefore in spite of its simplicity, according to some mathematical criteria it may fall into the "complex" category. In the present paper we reserve however the term "complex" for dynamics that are truly complicated to the degree that they do not admit simple verbal or mathematical description or understanding).  

The homogenous stock market described above exhibits booms and crashes. 
However, the homogeneity of investors leads to unrealistic periodicity.
As shown below, when there is more than one investor species the 
dynamics becomes much more complex and realistic.
  
\bigskip 
{\it Figure 2 : The Fourier transform of the price in a market 
with
 one species with extrapolation range $k=10$.
 The market contained 10000 traders that had initially 
equal wealth
 invested half in stock and half in bonds.}
  \bigskip 
{\bf 3. Realistic Features in LLS with Many Species } 

Our numerical experiments within the LLS framework have found that 
already a small 
number of trader
species (characterized by different extrapolation ranges $k$) 
leads qualitatively to many of
the empirically observed market phenomena. 

In reality, we would expect not just a few trader types, but rather an 
entire spectrum of investors. 
 When the full spectrum of different trader species
(fundamentalists and various other types - see [1] for 
the detailed operational definition) is considered it turns out 
that "more is different" [5]: the 
price dynamics becomes realistic: 
booms and crashes are not periodic or predictable, and they are also 
less frequent and dramatic. At the same time, we still obtain many of
 the usual market anomalies described by the experimental studies 
(however in the limit of infinite times or infinite number of 
investors, the dynamics may revert to predictable patterns 
[3]). 

We list below a few such realistic features: 

{\bf Return Autocorrelations: Momentum and Mean-Reversion} 

In the heterogeneous population LLS model trends are generated by the 
same positive  feedback 
mechanism that generated cycles in the homogeneous case (section 2): 
high (low) returns  tend to make the 
extrapolating investors more (less) aggressive, this generates more 
high (low)  returns, etc. 

The difference between the two cases is that in the heterogeneous 
case there is
 a very complicated 
interaction between all the different investor
species and as a result there are no distinct regular 
cycles but rather, smoother and more irregular trends.  
There is no single cycle length - the dynamics is 
a combination of many different cycles corresponding to the many 
extrapolation ranges $k$.  This makes 
the autocorrelation pattern also smoother and more continuous.  
The return autocorrelations in the 
heterogeneous LLS model conform to the empirical findings: 
In the short run the autocorrelation is positive - 
this is the empirically documented 
phenomenon known as momentum: 
high returns during a trading quarter tend to be followed by more 
high returns in the following months, 
(and low returns tend to be followed by more low returns). 
In the longer run the autocorrelation is negative 
(after a few years of boom, one
usually experiences a few "dry" years), 
which is known as mean-reversion. 
For even longer lags the autocorrelation 
eventually tends to zero [1]. The short run momentum, 
longer run mean-reversion, 
and eventual diminishing autocorrelation creates 
the general  "U-shape" that is found in empirical studies [7].

{\bf Excess Volatility} 

In markets with a large fundamentalist population 
(see [1] for  their detailed operative definition in the LLS model),
the price level is generally determined by the fundamental value of the 
stock.
  However, the market 
extrapolating investors occasionally induce temporary departures of the 
price away from the 
fundamental value. These temporary departures from the fundamental value
 make the price more 
volatile than the fundamental value. 

 Following Shiller's [8] methodology  we measured the 
standard deviations of the detrended price and 
fundamental value. Averaging 
over $100$ independent simulations we found [1] respectively  
$27.1$  and  $ 19.2$, which is an excess 
volatility of $41 \% $. 

{\bf Heavy Trading Volume } 

In an LLS market with both fundamentalists and market extrapolating 
investors (over various $k$ 
ranges), shares change hands continuously between the various groups: 

When a "boom" starts, the extrapolating investors observe higher ex-post
 returns and become 
more optimistic, while the fundamentalists view the stock as becoming 
overpriced and 
become more pessimistic. Thus, at this stage the market extrapolators 
buy most of the shares 
from the fundamentalists. 

When the stock crashes, the opposite is true: the extrapolators are 
very pessimistic, but the 
fundamentalists buy the stock once it falls below the fundamental 
value. 
Thus, there is substantial trading volume in this market. 
The average trading volume in a typical LLS 
simulation was about $1,000$ shares per period, 
or about $10 \%$ of the total outstanding shares. 

{\bf Volume is Positively Correlated with Absolute Returns } 

The typical scenario in an LLS run is that when a positive trend is 
induced by the extrapolating 
investors, the opinions of the fundamentalists and the extrapolating 
investors change in opposite 
directions: 

- The extrapolating investors see a trend of  rising prices as a 
positive indication about the 
future return distribution, while 

- The fundamentalists believe that the higher the price level is 
(the more overpriced the stock is),  the harder it will eventually fall. 

The exact opposite holds for a trend of falling prices. 
Thus, price trends are typically interpreted 
differently by the two investor types, and therefore induce heavy
 trading volume. The more 
pronounced the trend (large price changes), the more likely it is
 to lead to heavy volume. 

In order to 
verify this relationship quantitatively we regressed volume $V(t)$ 
on the absolute rates of 
return $r(t)$  for $100$ independent simulations. We run the 
regressions: 
$$ V(t) = a + b |r(t)| + random(t) $$ 
We found an average value of $870$ for $b$ with an average t-value of $5.0$.
Similar results were obtained for 
time lagged-returns. 

  \bigskip 
   {\bf 4. Predation, Competition and Symbiosis between Trader Species} 

In section 2 it was explained that a homogenous population of traders 
that extrapolate the last $k$ returns 
leads to cycles of booms and crashes of period $2k + O(1)$. 
When there are two species with 
extrapolation ranges $k_1$ and respectively $k_2$, 
we observe sharp irregular transitions between eras where 
one
 species dominates (cycles of period $2 k_1+ O(1)$)
 and market eras where the other species 
dominates (cycles of period $2 k_2 + O(1)$).
 When the number of trader species is three, there
are dramatic qualitative changes: generically, the dynamics becomes 
complex. We show that 
complexity is an intrinsic property of the stock market. 
This suggests an alternative explanation to the widely 
accepted but empirically questionable random walk hypothesis.
We discuss below some of the market ecologies possible with only two 
species of traders. Of course the picture becomes more complex later,
when 3 or more species are introduced. 

{\bf Market Ecologies with Two Trader Species } 

When there are two trader species with different extrapolation spans 
it turns out that the nature of the 
dynamics is determined by the ratio of the extrapolation spans of the
 two species. In [9] 
we performed a qualitative theoretical analysis of this phenomenon 
and supported it by microscopic 
simulations. We showed that in market eras in which one species 
(of extrapolation range $k_0$) dictates 
the dynamics (i.e. boom-crash cycles have periods of length 
$2 k_0+ O(1)$) the second species (with 
extrapolation range $k$)  has generically the following performance: 

A : If $ k_0 < k  < 2 k_0$   then $k$ is performing very poorly (looses money) 

B : If $ 2n k_0 < k < (2n + 1) k_0$  
(with $n$ natural number), then $k$ is doing relatively well 

C : If $ (2n + 1) k_0 < k < 2n k_0 , n > 1,$  then $k$ does better than in A but worse than in B 

D : If $ k < k_0 $,   then $k$ is doing well . 

These facts turned out sufficient to understand the main 3 cases that
 a 2-species ecology can display: 

{\bf Case 1: predator - prey dynamics} 

If one considers one species with an extrapolation range $k_1= 10$ 
and a second species with an 
extrapolation range $k_2= 14$ it turns out that the 
resulting ecology dynamics  
is a predator-prey one. 
In fact the LLS market dynamics leads in this case to the extinction 
(total impoverishment) of the $k_1$ 
species: after some time the entire wealth on the market belongs to the
 species $k_1=10$ (Fig. 3). As a 
consequence, the market price presents clear cycles of booms and 
crashes of periodicity clustered around $24= 2 * 10+ O(1)$. 

This is easily understood since according to the property A above the 
$k_2=14$ population is performing 
poorly when the $k_1$ dictates the market periodicity while the 
population $10$ is performing well 
according D in the hypothetical periods when $k_2=14$ dictates the 
market periodicity. 
This is only an example of a large class of parameters 
that lead to 
 predator-prey systems 
and which may result in the total extinction of one of the
species.
\bigskip

{\it Figure 3 : Fraction of the wealth that the species $k_1=10$
 possesses in Case 1. The traders in the market belonged to
 2 species consisting each of 5000 traders.
 Each trader owed at the beginning 5000
 dollars in cache and 5000 shares (worth each 1.4 dolars).}
\bigskip

{\bf Case 2: competitive species } 

 If one chooses $k_1=10, \  k_2=26$, the species with extrapolation 
range $26$ gains during the periods when 
the species $k_1= 10$ dominates (property B) but species $k_1=10$ gains 
when the species 
$k_2= 26$ dominates 
(property D). It is therefore reasonable that one species can not 
dominate the other indefinitely. Indeed, 
a look at the fraction of the wealth held by the species with 
extrapolation range 
$k_1=10$ reveals 
alternating eras of dominance (Figure 4). This is also reflected in 
the alternance between price cycles 
($\sim 56$) corresponding to $k_2=26$ and price cycles ($\sim 24$) 
corresponding to $k_1=10$. 
Clearly this alternance between the 2 species corresponds to a 
classical competitive ecology, in 
which two competing species take turns in dominating the ecology. 
Note however that most of the time it is the population $k_2$
which dominates the wealth. 
This seems to be a generic tendency in the long runs limit.

\bigskip

{\it Figure 4: Fraction of the wealth that the species $k_1=10$
 possesses in Case 2.
 The initial conditions were similar to Figure 3.} 
\bigskip

{\bf Case 3 symbiotic species } 

In the case $k_1=10, \  k_2=36$, similarly to the $10-26$ market, the 
investors with extrapolation range
$k_2=36$ 
are doing better than those with extrapolation range $k_1=10$ when 
$k_1=10$ dictates the dynamics 
(cf. property C). On the other hand  $k_1=10$ are doing better when the 
species $k_2=36$ dictates the dynamics 
(cf. D). Hence, we may speculate that again we will find alternating 
eras of dominance. Figure 5 
shows that this is not the case. The difference between this case and 
the $10-26$ 
case is that here the 
market remains stuck in a "metastable" state: 
the extrapolation range $36$ population never gains enough 
wealth to dictate long cycles. Thus, the system remains in a state of  
{\it symbiosis} throughout 
the run: the price cycles correspond to the short species extrapolation 
range span 
$k_1=10$ while $70-80 \%$ 
of the wealth stays with the long extrapolation span species $k_2$. 

For very long $k_2$ extrapolation ranges, 
the share of the total wealth detained by 
$k_2$ can be 
even larger (approaching unity). 
\bigskip
{\it Figure 5 -Fraction of the wealth that the species $k_1=10$
 possesses in case 3.} 
  \bigskip

In conclusion [9] has uncovered a quite lively ecology of the 
traders populations in the LLS model and 

{\it 
"observed phenomena ranging from complete dominance of one population 
to 
alternating eras of domination and to symbiosis. . . . 
Our results suggest that complexity is an intrinsic property of the 
stock  market. The dynamic 
and complex behavior of the market need not be explained as an 
effect of external random 
information. It is a natural property of the market, emerging from 
the strong nonlinear 
interaction between the different investor subgroups of the 
market . . ."} 

The main source of endogenous dynamics in the LLS model turns out to 
be the feedback between the market price fluctuations and the wealth 
of the investors belonging to various species:
 
 - On one hand the wealth of the investors determines their influence on the price changes (at the short range): e.g. the richest determine the periodicity of the boom-crash cycles.

 - On the other hand, the variations in the price determine changes in the distribution of wealth, which iterated over longer time intervals, result in changes in the market price cycle periodicity regime.  

The entire cycle of rise and fall 
of a given species can be schematically described as: 
$\rightarrow$ 
   The species has by chance a (momentary) winning strategy 
$\rightarrow$ 
   Investors belonging to the species gain wealth 
$\rightarrow$ 
   Overall wealth of the investors belonging to the species  increases 
$\rightarrow$ 
   Bids of investors belonging to the species  become large 
$\rightarrow$ 
   Investor bids influence the market price adversely (self-defeating) 
$\rightarrow$ 
   Trading of  investors belonging to the species becomes inefficient 
$\rightarrow$ 
   Investors lose money 
$\rightarrow$ 
   Investors belonging to the species become poor 
$\rightarrow$ 
   Species wealth and market relevance vanish
$\rightarrow$ 
   Other species with different strategies become winners 
$\rightarrow$ 
   Cycle re-starts (with the new winning strategies). 

A few comments are in order: 

 {\bf 1.} the concept of efficient strategy is only a temporary one as it 
depends crucially of the state 
of the market: by its very efficiency at a certain moment, a strategy 
prepares the seeds of its  failure in the future. 

 {\bf 2.} the biological and cognitive analogies are useful but their limits 
should be understood: 

 - in biology, the species selection mechanism is based on the 
disappearance of the inefficient 
individuals. 

 - in the learning adaptive agents' case, the individuals discard 
loosing strategies for new ones. 

In the LLS market framework, while it is possible to include the above 
effects, they are not necessary: 
the strategies selection takes place automatically by their carriers 
(traders belonging to the species) losing or gaining: 
for the market to be efficient, no {\it a priori} intelligence 
nor explicit criteria for the evaluation and comparison of market 
performance are required: 
just the natural (Adam Smith's "invisible hand") market mechanisms. 

 {\bf 3.} While the adverse influence on the market price 
implied by the large orders coming from rich agents' 
leads automatically to inefficiency in their operations
(except for rich agents which follow a buy-and-hold strategy and therefore do not influence (adversely) the market), 
the mere lack of market influence due to 
poverty does not guarantee a winning strategy. 
It is necessary therefore that there are enough 
strategies and enough agents in the market for insuring its 
efficiency. 

{\bf Three Investor Species } 

One might suspect that the three species dynamics is a natural 
extension of the two species dynamics. 
Instead of alternating between two cycle lengths the system may 
just alternate between the three 
possible states of dominance. Figures 6-7 shows that this is not 
at all the case. These figures depict a 
typical part of the dynamics of a three species market, with 
extrapolation spans of  10, 141 and 256 
respectively. With the introduction of a third species the system has 
underwent a qualitative change: 
there is no specific cycle length describing the time series. 
Instead, we see a mixture of different time 
scales: the system has become complex. Prediction becomes very 
difficult, and in this sense the 
market is much more realistic. Figure 6 shows the power struggle 
between the three species while the 
Figure 7 depicts the Fourier transform of the price evolution during 
this run. 

Although the dynamics is complex, it is clear from Figures 6 and 7 that
 there is an underlying structure, 
which perhaps may be analyzed by the properties A, B, C, D and their 
generalizations. For instance it would appear from the Figure 6 that 141 and 256 take turns in dominating while 10 has a chance to a non-vanishing wealth share only occasionally in the transition intervals between 141 and 256 dominated eras. The dynamics 
generated by only three investor species can be extremely complex, even 
without any external random 
influences. 
\bigskip
{\it Figure 6:  The species wealths in a market with 3 species of
 extrapolation ranges of respectively 10, 141 and 256 days.
 Initially the 3 species possessed equal wealth
 distributed equally between stock and bond.
  Each species consisted of 1000 traders. } 
\bigskip
{\it Figure 7: Fourier transform of the stock price
 time evolution in the market described in Fig 6. } 
  \bigskip
\bigskip 
{\bf 5. Generalized Lotka-Volterra models 
for markets with multiple species} 

Inspired by the above facts we devised an effective dynamics that 
stylized the features uncovered in the 
LLS model and extended them to a more generic framework. Instead 
of following in detail the way the 
market price influence each species and individual $i$ , 
we assumed that this influence can be represented through multiplying  
their wealth $w_i (t)$ by stochastic 
multiplicative factors $\lambda_i  (t)$. 
This is natural in the LLS model in which the investments of the 
individuals (and consequently their returns) 
are fractions of their wealth 
(as implied by the constant relative risk aversion utility functions). 
The stochastic proportionality between personal returns and personal 
wealth is consistent with the real data that show that 
the (annual) individual income distribution is proportional to 
the individual wealth distribution [10]. 

We proposed [11] therefore a model including the above 
stochastic autocatalytic 
properties of the capital as well as the cooperative, 
diffusive and competitive/ predatory interactions between the
species. 
The resulting model turned out to be a straightforward generalization 
[11] of  the  Lotka-Volterra 
system (discrete logistic equation) well known previously in population 
biology: 
$$ 
w_i (t+1) = \lambda_i(t) w_i (t) +  \sum_k a_k w_k (t)
-   w_i  \sum_k b_k w_k (t) 
$$ 
Where the sum is over all the $N$ traders participating in the market.
There are a few other (mutually non-exclusive) possible interpretations 
for $w_i$ in addition to the individual
wealth: 
the wealth associated with a particular investing strategy, the 
capitalization associated with a particular 
company/industry, or the number of investors following a common trend 
(herd). 

Particularly interesting cases were studied subsequently: 

1) The linear case where the total wealth diverges to larger and 
larger values (inflation, production): 
$$ 
w_i (t+1) = \lambda_i  (t) w_i (t) +   \sum_k  a_k w_k (t) 
$$ 

2) The case in which the binary interactions between individuals are 
expressible in terms of interactions with the total wealth 
$W =\sum_k w_k $: 
$$  w_i (t+1) = \lambda_i  (t) w_i (t) +   a W (t) 
-  b w_i  W (t) $$ 

3) The case in which individual wealth is bounded from below
by a certain fraction $c$ of the average wealth 
${\bar w} = W/N$ [12]: 
$$ 
w_i (t+1) = \lambda_i  (t) w_i (t) 
$$ 
except if 
$$ 
\lambda_i  (t) w_i (t) < c \bar w 
$$ 
when 
$$ 
w_i (t+1) = c \bar w 
$$ 

4) The case of the random multiplicative wealth dynamics 
$$ 
w_i (t+1) = \lambda_i  (t) w_i (t) 
$$ 
with variable number of traders [13]: 

 - traders which fall below a certain fixed minimal wealth 
$w_{min}$ drop from the market 

 - a number of traders proportional to the total wealth increase: 
$$\Delta N = c (W (t+1) - W(t))/ w_{min}$$ 
join the market at each time step
i.e. the average wealth remains constant
in this process: 

$$\bar w =  w_{min} / c$$.

One assumes that each of the new traders
brings an initial investment equal to $w_{min}$ 
which means that the total amount of added wealth
is $$\Delta N w_{min} = c (W (t+1) - W(t))$$
 i.e. a fraction  $c $ of the total 
wealth increase $(W (t+1) - W(t))$.

In most of these systems were assumed asynchronous: at each time step, 
only one (randomly chosen) $w_i$ was updated. 
Very striking generic results can be obtained with all these models in 
certain relevant regimes. We limit ourselves
below to the universal scaling properties (power laws). 

  \bigskip 
   {\bf 6. Pareto Law in LLS and Lotka-Volterra  models } 

The efficient market hypothesis and the Pareto law are some of the 
most striking and basic concepts in economic thinking. 
It is therefore very important that our models above succeed to 
connect them in a very essential way. 

Let us discuss this in more detail. More than a hundred years ago, 
Pareto [14] discovered that the number of 
individuals with wealth (or incomes) with a certain value  
$w$ is proportional to $w^{-1-\alpha}$. 
This later became known as the Pareto Law.
The LLS model 
treats the individual investor wealth as a crucial quantity, 
and it views its feedback relation to the 
market dynamics as the main source driving the endogenous dynamics 
of the market. 

It turns out that in the conditions in which the participants in the
market do not have a systematic advantage one over the other 
(which is in fact expected in an efficient market), a dynamics of 
the LLS type leads always to a Pareto law. The actual 
value of the exponent $\alpha$ depends on the particular parameters 
used in the model. Mainly, as 
explained below $\alpha$ is influenced by the 
social security policy.  If one does do not allow any individual 
to become poorer than a certain fraction c of the current average 
wealth then, for a wide range of conditions
$\alpha = 1/(1-c)$. This is confirmed in Figure 8 which plots the 
wealth distribution in the LLS model with $k=3$
and $c=0.2$ (and $U(W) = \ln W$). 
\bigskip
{\it Figure 8: The wealth distribution of the investors in an LLS
 model with a poverty line of  $c = 20 \%$ of the average wealth.
 One a double logarithmic
 scale one obtains a straight line with slope 2.2
 corresponding to an $\alpha$ of 1.2.

 The market consisted of 10000 traders and the measurement
 was performed as a "snapshot" after 1 000 000 "thermalization" 
market steps.
 Initially all the traders had equal wealth ($\$$1000) equally
 distributed between bond and stock.}
\bigskip

In fact it has been proven [11-13] theoretically that any of the 
effective dynamics of the type 1-4 with $\lambda_i$
distribution independent on $i$ or $w_i$ leads always to a power law 
Pareto distribution. 

In a wide range of models, the generic rule is that 

{\bf $\alpha$ = 1/(1-c)} 

where $c$ is essentially 
the market global impact factor [13]: 

{\bf c= exogenous new capital ADDED to the market /  
increase in stock capitalization due to market price increase}

Since the increase in the capitalization is
the increase in wealth that the owners incurr
upon their investment of
new capital, the ratio can be also interpreted 
as the long range market return factor 

{\bf c = 1/ (long range market return factor)}.
 
Let us explain in short how such results were obtained [15].
 The crucial observation is that for large $w_i$ values,
the non-stationary 
multiplicative system of interacting $w_i$'s 
is formally equivalent to a statistical mechanical (additive) system 
in thermal equilibrium  when expressed in terms of the
variables $u_i (t) = \ln {w_i  / \bar w (t)}$. 
For instance the system 3)  is mapped 
into a system of particles diffusing in an energy potential field $u$
with a ground level $u_0 = \ln c$. 
In thermal equilibrium, all such systems 
(independent on the details of the 
interactions between their particles)  
have an universal probability distribution discovered by 
Boltzmann more than 100 years ago: 

$$ 
P(u) \sim exp(- \alpha u) 
$$ 
When re-expressed in terms of the original $w_i$ variables this gives a 
Pareto power law distribution: 
$$ 
P(w) \sim w^{-1-\alpha} 
$$ 
The exponent $\alpha$  can be estimated from
the integrals representing the total wealth and the
total number of traders [12]. 
For instance, in the models 3-4, in the limit of 
$N \rightarrow \infty$ , the result is:
$$\alpha = 1/(1- w_{min} / {\bar w}) $$ 
i.e.
$$\alpha = 1/(1-c) $$ 

Thus, the Pareto law is the exact analogue of the Boltzmann 
law for stochastic systems that are multiplicative rather 
then additive.

For finite $N$ the $\alpha$ is given by the implicit transcendental
equation [16]:

$$ N = [ ((1- (N/c)^\alpha )    / \alpha ] /
       [ ((1- (N/c)^{\alpha -1})/(\alpha - 1) ]
$$
which for $ N \ll e^{1/c}$ gives approximately:
$$ \alpha \sim {\ln N } / ({\ln (N/c)}) < 1$$  
which incidentally means that in this regime all the wealth belongs
to only a few individuals. 

In the system 1 defined above, in the appropriate thermodynamic limit
The analog result is: 
$$\alpha = 1/(1-c) $$ 
with
$$ c \sim 2a/(< \lambda  > + \sigma^2 / 2 )$$ 
 where $\sigma$ is the standard deviation of $\lambda$ 

And in the case 2: 
$$ c \sim 2a / (\sigma^2 + a^2) $$ 
independent on $b$ and $<\lambda  >$. 

  \bigskip 
   {\bf 7. Market Efficiency, Pareto Law and Thermal Equilibrium} 

The formal equivalence between the non-stationary systems of 
interacting $w_i$'s  and the equilibrium statistical
mechanics systems governed by the universal Boltzmann distribution 
has far reaching implications: 
it  relates the Pareto distribution to the 
efficient market hypothesis: 
In order to obtain a Pareto power law wealth distribution it is 
necessary and sufficient that the returns of all the
strategies practiced in the market are stochastically the same,
i.e. there are no investors that can obtain "abnormal" returns. 

Therefore, the presence of a Pareto wealth distribution is a measure of 
the market efficiency in analogy to the
Boltzmann distribution whose presence is a measure to thermal 
equilibrium. 
Indeed physical systems which are {\bf not} in thermal equilibrium 
(e.g. are forced by some external field - say by laser pumping) 
do {\bf not} fulfill the Boltzmann law. 
Similarly, markets that are not efficient 
(e.g. when some groups of investors make systematically more profit 
than others) do not yield power laws (see Fig 9). 
Optimal market and power laws are the short time and long time faces of 
the same medal/phenomenon. 

This analogy is consistent with the interpretation of market 
efficiency as analog to the Second law of Thermodynamics: 

- one can extract energy (only) from systems that are not in thermal 
equilibrium

- one can extract wealth (only) from markets that are not efficient. 
 \bigskip 
- by extracting energy from a non-equilibrium thermal system one gets 
it closer to an equilibrium one. 

- by extracting wealth from a non-efficient market one brings 
it closer to an efficient one 
 \bigskip 
-in the process of approaching thermal equilibrium, 
one also approaches the Boltzmann energy distribution 

- in the process of approaching the efficient market one also 
approaches the Pareto wealth distribution. 
 \bigskip 
-by having additional knowledge on a thermodynamic system state 
one can extract additional energy (e.g.
Maxwell demons gedanken experiment)

-by having additional knowledge on a financial system one can extract 
additional wealth.

\bigskip 

This double analogy 

\centerline{
thermodynamic equilibrium $\sim$ efficient market }

\centerline{
Boltzmann law $\sim$ Pareto law }

holds in the details of their  microscopic origins:
\bigskip

-  the convergence to statistical mechanics equilibrium  depends on  
the balance of the probability  flow entering
and exiting each energy level. 
This is usually insured microscopically by the fact that the a priori 
probability for a molecule to gain or loose an
energy quanta in a collision is the same for any energy level with the 
exception of the collisions including
molecules in the ground state which can only receive (but not give) 
energy. 

- in the stochastic models 1-4,  the  convergence of the wealth to the 
power-law is insured by the balance of
flow of investors from one level of [log (relative wealth)] to another. 
At the individual level, this is enforced by all the individuals having 
the same (relative) returns probability distribution 
(except for the individuals possessing the lowest allowed wealth). 
If this condition is not fulfilled, one does not get a wealth 
distribution power law. 

These facts should guide us in the practical runs in establishing
which combinations of strategies (or the strategy selection strategies) 
are producing a realistic market "in the Pareto sense". 
In Figure 9 one sees the wealth distribution in a model in which there 
are 2 trader species with slightly different distributions of $\lambda$.
 One sees that even a small violation of the $\lambda$ uniformity 
leads to significant departures from the Pareto law which are 
inconsistent with the historical experimental facts. 
The absence of such departures in real 
life is a strong indication of the market
efficiency in the weak stochastic sense
(that all investors have stochastically the same 
relative returns distribution). 
\bigskip
{\it Figure 9: Wealth distribution for 2 investor species with
 different return distributions. Model 3 was used with a lower
 wealth bound of $c= 20 \%$. 

 $\lambda$ is randomly drawn. 
For the first
 species $\lambda$ is 1.10 or 0.95 with equal probability. 
For the second
 "more talented" species $\lambda$ is 1.11 or 0.96 with equal
 probability.  The 2 species were each composed of 10000 traders
with initially equal wealth (1000 dollars each). 

The measurement
 of the wealth distribution
 was performed after a "thermalization period" of 100 000  
wealth updatings.} 
\bigskip

   {\bf 8. Leptokurtic Market Returns in LLS } 

It has been long known that the distribution of stock returns is 
leptokurtic or "fat-tailed". Furthermore, a specific functional 
form has been suggested for the short-term return distribution
(at least in a certain finite range) - the Levy distribution 
[17]. This feature is present in the LLS
model, and is directly related to the Pareto distribution of wealth. 

The central limit theorem insures that in a wide range of conditions 
the distance reached by a random walk of $t$
steps of average squared size $s^2$ is a 
Gaussian with standard deviation $s \sqrt t$. 
Suppose that at time $t=0$ one has  $N$ positive numbers 
$w_i (0); \ i=1,....,N$ of order 1 and sum $W(0)$. 
Suppose that at each time step one of the numbers varies 
(increases or decreases) by a fraction $s_i (t)  \ll 1$
extracted from a random distribution with average squared  $s^2$ 
(and $0$ mean). 
What will be the probability distribution of the sum  
$W(t)$ after $t$ steps? 
According to the central limit theorem this would be the Gaussian
$$ P(W,t) = 1/({\sqrt{ 2\pi t s^2}}) e^{-(W(t)-W(0))^2 / 2 t s^2} $$ 
since it consists of $t$ steps of average squared size $s^2$. 

If one interprets $w_i (t)$ as the value of the stocks owned by the
 trader $i$ at time $t$, then $W(t)= \sum_i w_i$ is the total
market value of the stock and therefore $ (W(t)-W(0))/ W(0)$ is the 
relative stock return for the time interval $t$. 

One sees that if the central limit theorem would hold, one would 
predict a Gaussian stock returns distribution. 
This is in fact the case for real stocks and time intervals longer 
than a few weeks. 
For significantly shorter times $t$ however, the distribution of 
returns is very different from a Gaussian. Even
though the exact shape of the returns distribution is not yet 
established experimentally, it is generally agreed that in certain
ranges (typically "in the tails"- i.e. for large $w_i$ values) 
it fits better a power law rather than a Gaussian. 

Such a situation can in principle be explained by the following 
scenario: 

Suppose that at time $t=0$ one has an arbitrarily large number 
of positive numbers $w_i (0)$. 
Suppose moreover that the probability distribution for 
the sizes of 
$w_i (0)$ is 
$$ 
P(w) \sim w^{-1-\alpha} 
$$ 
Suppose that at each time step one of the $w_i $'s varies 
(increases or decreases) by a fraction $s_i (t)  \ll 1$ of
average squared size  $s^2$. What will be the probability 
distribution of the variation of the $w_i $'s sum 
$W(t) - W(0)$ after $t$ steps? 

One is tempted to think that the correct answer is given by 
$$ P(W,t) = 1/({\sqrt {2\pi t s^2}}) e^{-(W(t)-W(0))^2 / 2 t s^2} $$ 
for some $s$. However this is wrong. 
Indeed, assuming such an $s$ exists would imply that the 
probability for the sum variation $W(t) - W(0)$ to be 10 after a time
$t= 1/(2 s^2)$ is: 
$$ P(W(t)=W(0)+10,t= 1/ (2 s^2) ) \sim  e^{-10^2} \sim 10^{-32} $$ 
while in reality a lower bound for the probability of getting $W(t)-W(0)=10$ in  just one step it is obviously that given by
$$ 
P(w) \sim w^{-1-\alpha} 
$$ 
I.e. $ P(W(t)=W(0)+10,t= 1/ (2 s^2) ) $ is at least of 
order 
$10^{-1-\alpha}$ 
which for $\alpha < 2$ means it is larger than $10^{-3}$ ! 

This coarse estimations highlights the difference between 
the Gaussian 
distributions and the distributions generated by random walks 
with power distributed step sizes (called Levy distributions 
[18,17]):
the presence of $w_i$'s of arbitrary size 
implied by a power law distribution
insures that the large returns distribution is dominated by the power 
law of the individual step sizes rather than the
combinatorics of the multiple events characterizing the Gaussian system. 

One sees now that the systems 1-4 (and consequently LLS)
are exactly of the type one needs to 
explain returns distribution power tails: 

- on one hand according to section 6, the models 1-4 
(and consequently LLS) insure a power distribution of $w_i$'s. 

- on the other hand, in the models 1-4 the variation of the stock 
index W(t) is the sum of the variations of the
individual $w_i (t)$'s. 

- these variations $w_i (t+1) - w_i (t) $ are stochastic 
fractions $s_i (t) = \lambda_i (t) -1$ of $w_i$ as above 
(the fact that $\lambda_i (t) -1$ has not $0$ mean is taken
care by working actually with $u_i = \ln (w_i / \bar w )$). 

Therefore, according to the argument above, the effective  
models 1-4, which reflect the stochastic
proportionality in LLS between individual wealth, individual 
investments and 
individual gains/losses predict that the price fluctuations 
in the LLS model will obey a Levy distribution (and in
particular fit a power in some range of the "tail"). 

There is a proviso for this argument to hold: the number of individual 
terms $N$ has to be larger than the 
number of time steps $t$. Otherwise the finite size of the sample 
of $w_i$'s will show up in the absence of sizes 
$w_i$ larger than a certain value. In fact for $t ~ N$ one 
recovers (slowly) the Gaussian distribution. 

In the LLS case, if the portfolio updatings are performed 
simultaneously by all the investors, the unit 
time step corresponds already to a time $t=N$. In order to 
verify the (truncated) Levy distribution 
and the power "tail" predictions, one has to look at the dynamics at a 
finer time scale. We therefore performed [12] LLS
runs in which at each time step only one trader  $i$ reconsiders 
its portfolio investment 
proportion $X(i)$. In such conditions, one expects to obtain 
a distribution which fits in a significant range a power
law (up to large $w_i$ values where the finite $N$ effects 
become important). 

This is in fact confirmed by the numerical experiments. 
While for the global updating steps one gets a Gaussian 
distribution,  for the trader-by-trader procedure
one obtains a truncated Levy distribution (Fig. 10). 
\bigskip
{\it Figure 10: The returns distribution in the LLS model in 
which only one trader re-evaluates his/her portfolio per unit 
time. $c= 0.2$, $ k= 3$, U= $\ln W$.

 The market contained 10 000 traders with initially
 equal wealth and portfolio composition
 (half in stock and half in bonds).

 The number of market returns  in intervals of 0.001
 were measured during 5 000 000 market steps 
(after an initial 1 000 000 equilibration period).} 
\bigskip

Note that in the central region of the short time returns 
(before the cut-off becomes relevant) the Levy
distribution is characterized by an $\alpha$ equal to the
exponent $\alpha$ of the traders' wealth distribution.

As explained in Section 6, in certain conditions
(e.g. model 4) one can interpret
$\alpha$ as 

{\bf $\alpha$ = 1/(1- 1/(long  term  market  return  factor))}.

Therefore the analysis above relates the 
{\bf stochastic distribution
of the short term returns} to the 
{\bf value of the long term returns} via the {\bf exponent of
the Pareto power law of individual incomes/wealths}.

Moreover the {\bf long term returns} are related 
(e.g. model 3)
via the value of the Pareto exponent $\alpha$ to the
ratio ($\bar w / w_{min}$)
between the average wealth/income and the 
{\bf lowest admissible wealth/income}:
the value $\alpha \sim 1.4$ implies 
(cf. models 3-4) for both
these quantities values of the order of 
$$1/c = \alpha /(\alpha -1 ) \sim 3.5.$$

Speculatively, one may try
to use the above relation in order to
explain the stability
of the Pareto constant $\alpha$ over the past century 
(and over the various countries and economies).

Indeed one may relate the implied value $3.5$
for both $\bar w / w_{min}$
and the long term market return to some basic 
biological invariant which is the average  number of
dependents / offsprings humans have: 

- if $w_{min}$ is the minimal amount necessary to keep
alive one person in a certain society (cost of life),
then the average income $\bar w$ will have to equal roughly
$w_{min}$ times the number of dependents the average 
household head has to
support.

- at the social level, the total effort/wealth that
one generation invests  in the economy has to result
in an economical growth capable to support a population larger
by a factor equal to the average number of descendents.

\bigskip 
{\bf Acknowledgement}
We thank T. Lux and D. Stauffer for very intensive and detailed 
correspondence on various simulation experiments using the LLS model.
\bigskip 

{\bf References} 
\parindent 0pt 

1. 
M. Levy, H. Levy, S. Solomon, 
{\it Microscopic Simulation of Financial Markets}, 
Academic Press, New York, 2000.

2. 
Levy, M., Levy H. and S. Solomon, "A Microscopic Model of the Stock Market : Cycles, Booms, and Crashes,"  Economics Letters, 45, 1994.

Moss de Oliveira S., H. de  Oliveira and D.Stauffer, Evolution, Money, War and Computers, B.G. Teubner Stuttgart-Leipzig 1999. 

Solomon, S. [1995], The microscopic representation of complex macroscopic phenomena, Annual Reviews of Computational Physics II, 243-294, D. Stauffer (editor), World Scientific 1995.

Solomon S. Behaviorally realistic simulations of stock markets;
Traders with a soul, Computer Physics Communications 121-122 (1999) 161

3.
Hellthaler T., "The Influence of Investor Number on a Microscopic Market", Int. J. Mod. Phys.C 6, 1995

Kohl R., "The Influence of the Number of Different Stocks on the Levy, Levy Solomon Model," Int. J. Mod. Phys. C 8, 1997.

4.  
Levy, M., Levy H. and S. Solomon, "Simulation of the Stock Market: The Effects of Microscopic Diversity," Journal de Physique I,   5, 1995.

5. 
Anderson P. W., J. Arrow and D. Pines [1988], eds. The Economy as an Evolving Complex System (Redwood City, Calif.: Addison-Wesley, 1988);

6. 
Egenter E., T. Lux and D. Stauffer, "Finite Size Effects in Monte Carlo Simulations of Two Stock Market Models," Physica A 268 ,1999.

7. 
Fama, E., and K. French, "Permanent and Temporary Components of Stock Prices," Journal of Political Economy, 96, 1988.

8. 
Shiller, Robert J., "Do Stock Returns Move too Much to be Justified by Subsequent Changes in Dividends?" American Economic Review, 1981.

9. 
Levy M., Persky N., and Solomon S., 
"The Complex Dynamics of a Simple Stock Market Model,
" International Journal of High Speed Computing, 8, 1996.

$http://www.ge.infm.it/econophysics/papers/solomonpapers/stock\_ex\_model.ps.gz
$
see also

Farmer J.D., "Market Force, Ecology and Evolution," e-print adap-org/9812005. See also J.D. Farmer and S. Joshi, "Market Evolution Toward Marginal Efficiency" SFI report 1999.

10. 
Levy, M., and S. Solomon, "New Evidence for the Power Law Distribution of Wealth", Physica, A 242, 1997. 

Levy, M. "Are Rich People Smarter?" UCLA Working Paper 1997.

Levy, M. "Wealth Inequality and the Distribution of Stock Returns"
Hebrew University Working Paper 1999.     

11.
Solomon S. and M. Levy, "Spontaneous Scaling Emergence in Generic Stochastic Systems," International Journal of Modern Physics C , 7(5), 1996. 

S. Solomon,  Stochastic Lotka-Volterra systems of competing auto-catalytic agents lead generically to truncated Pareto power wealth distribution, truncated Levy distribution of market returns, clustered volatility, booms and crashes, In Computational Finance 97, Eds. A-P. N. Refenes, A.N. Burgess,  J.E. Moody, 
     (Kluwer Academic Publishers 1998)

O. Biham, O. Malcai, M. Levy and S. Solomon, Phys Rev E 58, 1352, (1998).

12.
Levy M., S. Solomon, "Power Laws are Logarithmic Boltzmann Laws" International Journal of Modern Physics C , 7 ( 4) 1996.

13. 
Aharon Blank and  Sorin Solomon 
cond-mat/0003240 Power Laws and Cities Population
  
14. 
V. Pareto, Cours d'Economique Politique, 2 (1897).

15. 
Sorin Solomon, cond-mat/9901250;
Generalized Lotka-Volterra (GLV) Models and Generic Emergence of
Scaling Laws in Stock Markets;
To appear in the proceedings of the 
"International Conference on Computer Simulations 
and the Social Sciences, Paris 2000" Hermes Science Publications.

16. 
O. Malcai, O. Biham and S. Solomon, 
Phys. Rev. E, 60, 1299, (1999).

17. 
B. B. Mandelbrot, Comptes Rendus 232, 1638 (1951)

H. A. Simon and C. P. Bonini, Amer. Econ. Rev. 607 (1958)

Mantegna, R. N., "Levy Walks and Enhanced Diffusion in the Milan Stock Exchange." Physica A, 179, 1991.

Mantegna, R. N., and Stanley, H. E.,  "Scaling Behavior in the Dynamics of an Economic Index." Nature, 376, 1995.

18. 
Paul Levy,
Theorie de l'Addition des Variables Aleatoires,
Gauthier-Villiers, Paris 1937.

19
Lotka, A.J., (editor) Elements of Physical Biology, Williams and Wilkins, Baltimore, 1925; 

V. Volterra [1926], Nature, 118, 558.

\bigskip 

\bigskip 

{\bf Figure Captions}
\medskip 

{\it Figure 1: The Flow Chart of the LLS market framework}
\bigskip 

{\it Figure 2 : The Fourier transform of the price in a market 
with
 one species with extrapolation range $k=10$.
 The market contained 10000 traders that had initially 
equal wealth
 invested half in stock and half in bonds.}
  \bigskip 

{\it Figure 3 : Fraction of the wealth that the species $k_1=10$
 possesses in Case 1. The traders in the market belonged to
 2 species consisting each of 5000 traders.
 Each trader owed at the beginning 5000
 dollars in cache and 5000 shares (worth each 1.4 dolars).}
\bigskip

{\it Figure 4: Fraction of the wealth that the species $k_1=10$
 possesses in Case 2.
 The initial conditions were similar to Figure 3.} 
\bigskip

{\it Figure 5 -Fraction of the wealth that the species $k_1=10$
 possesses in case 3.} 
  \bigskip

{\it Figure 6:  The species wealths in a market with 3 species of
 extrapolation ranges of respectively 10, 141 and 256 days.
 Initially the 3 species possessed equal wealth
 distributed equally between stock and bond.
  Each species consisted of 1000 traders. } 
\bigskip

{\it Figure 7: Fourier transform of the stock price
 time evolution in the market described in Fig 6. } 
  \bigskip

{\it Figure 8: The wealth distribution of the investors in an LLS
 model with a poverty line of  $c = 20 \%$ of the average wealth.
 One a double logarithmic
 scale one obtains a straight line with slope 2.2
 corresponding to an $\alpha$ of 1.2.

 The market consisted of 10000 traders and the measurement
 was performed as a "snapshot" after 1 000 000 "thermalization" 
market steps.
 Initially all the traders had equal wealth ($\$$1000) equally
 distributed between bond and stock.}
\bigskip

{\it Figure 9: Wealth distribution for 2 investor species with
 different return distributions. Model 3 was used with a lower
 wealth bound of $c= 20 \%$. 

 $\lambda$ is randomly drawn. 
For the first species $\lambda$ is 1.10 or 0.95 
with equal probability. 
For the second
 "more talented" species $\lambda$ is 1.11 or 0.96 with equal
 probability.  The 2 species were each composed of 10000 traders
with initially equal wealth (1000 dollars each). 

The measurement of the wealth distribution was performed after a 
"thermalization period" of 100 000  wealth updatings.} 
\bigskip

{\it Figure 10: The returns distribution in the LLS model in
 which only one trader re-evaluates his/her portfolio per unit 
time.
 $c= 0.2$, $ k= 3$, U= $\ln W$.

 The market contained 10 000 traders 
with initially
 equal wealth and portfolio composition
 (half in stock and half in bonds).

 The number of market returns  in intervals of 0.001
 were measured during 5 000 000 market steps 
(after an initial 1 000 000 equilibration period).} 
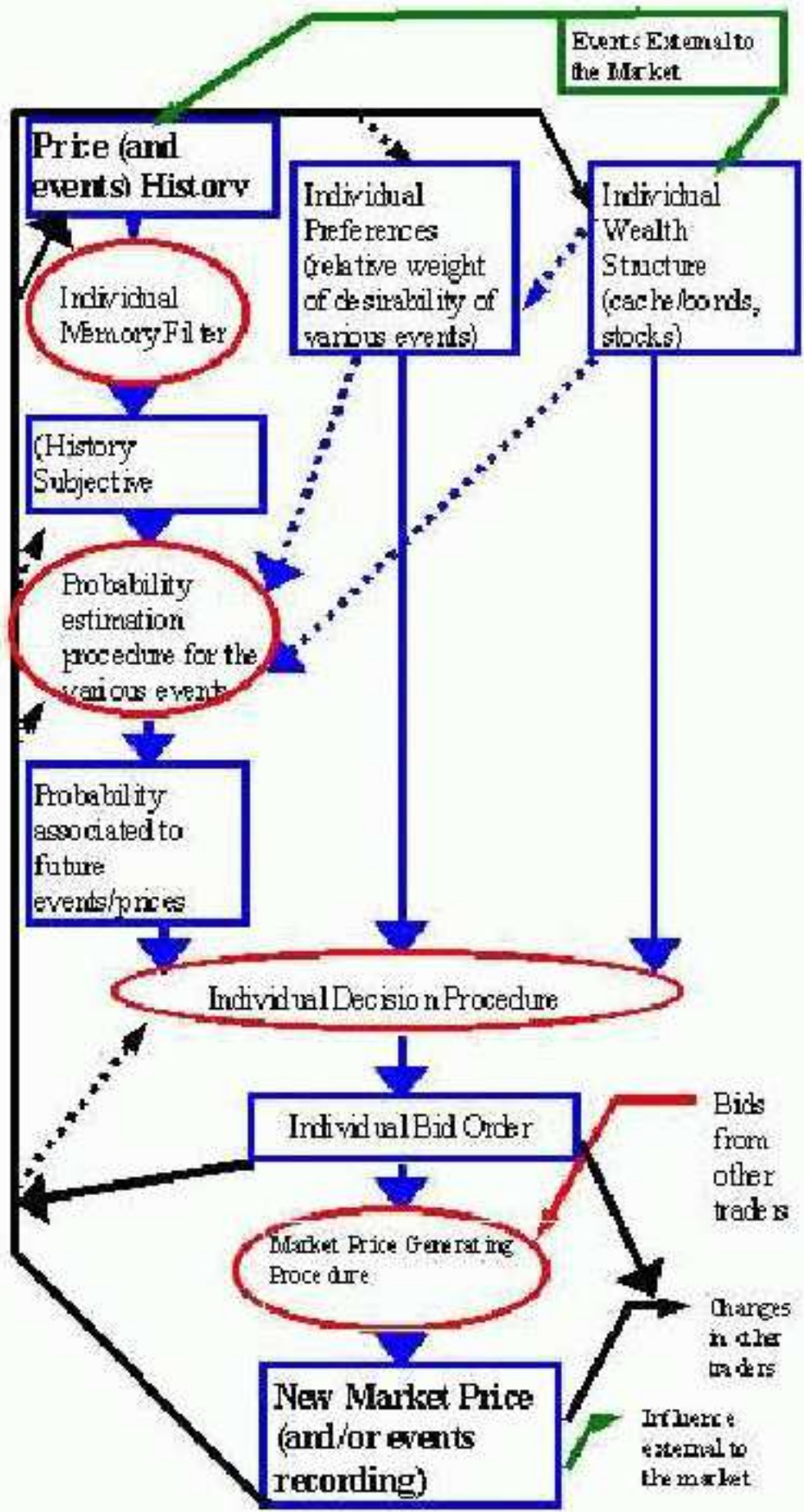
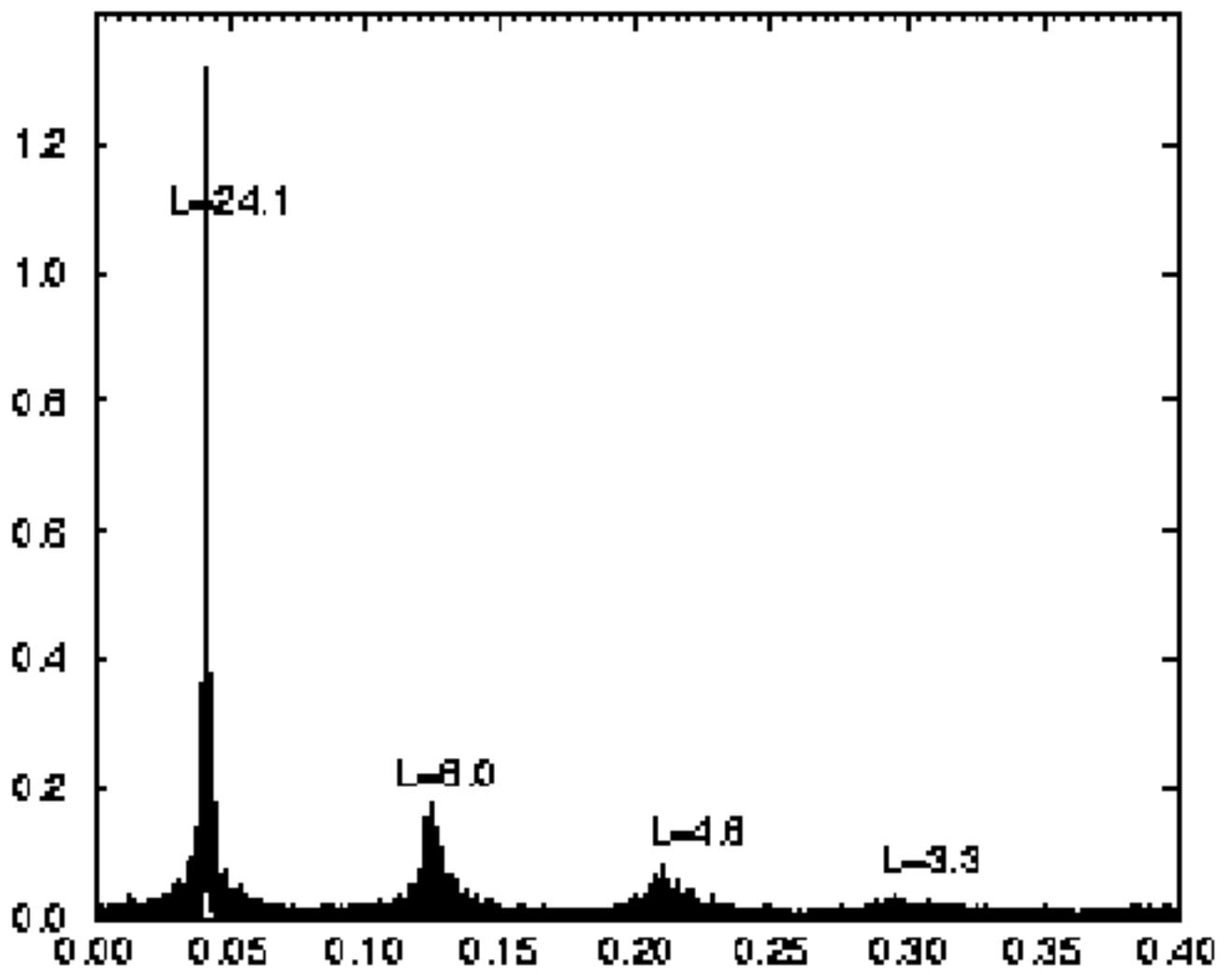
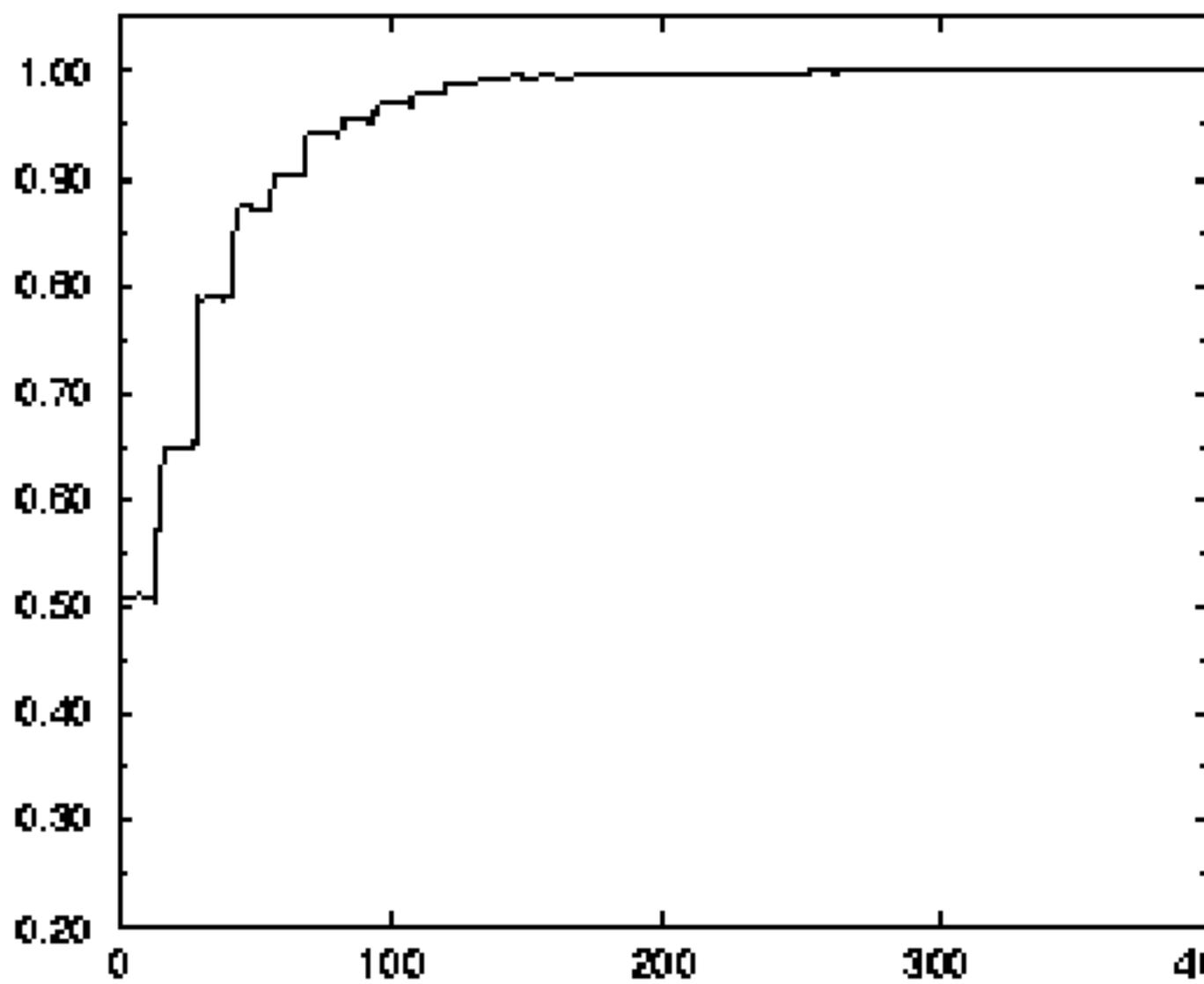
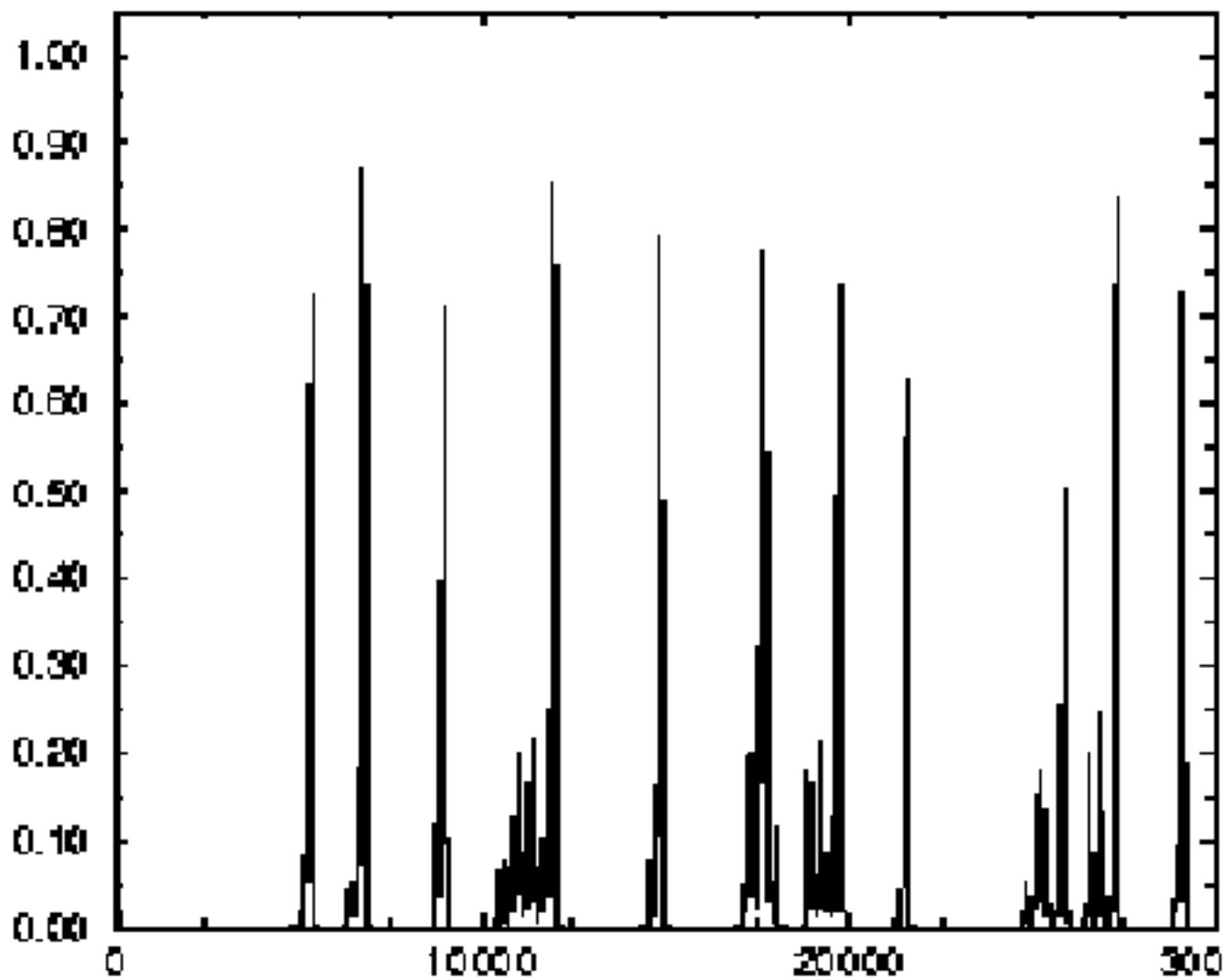
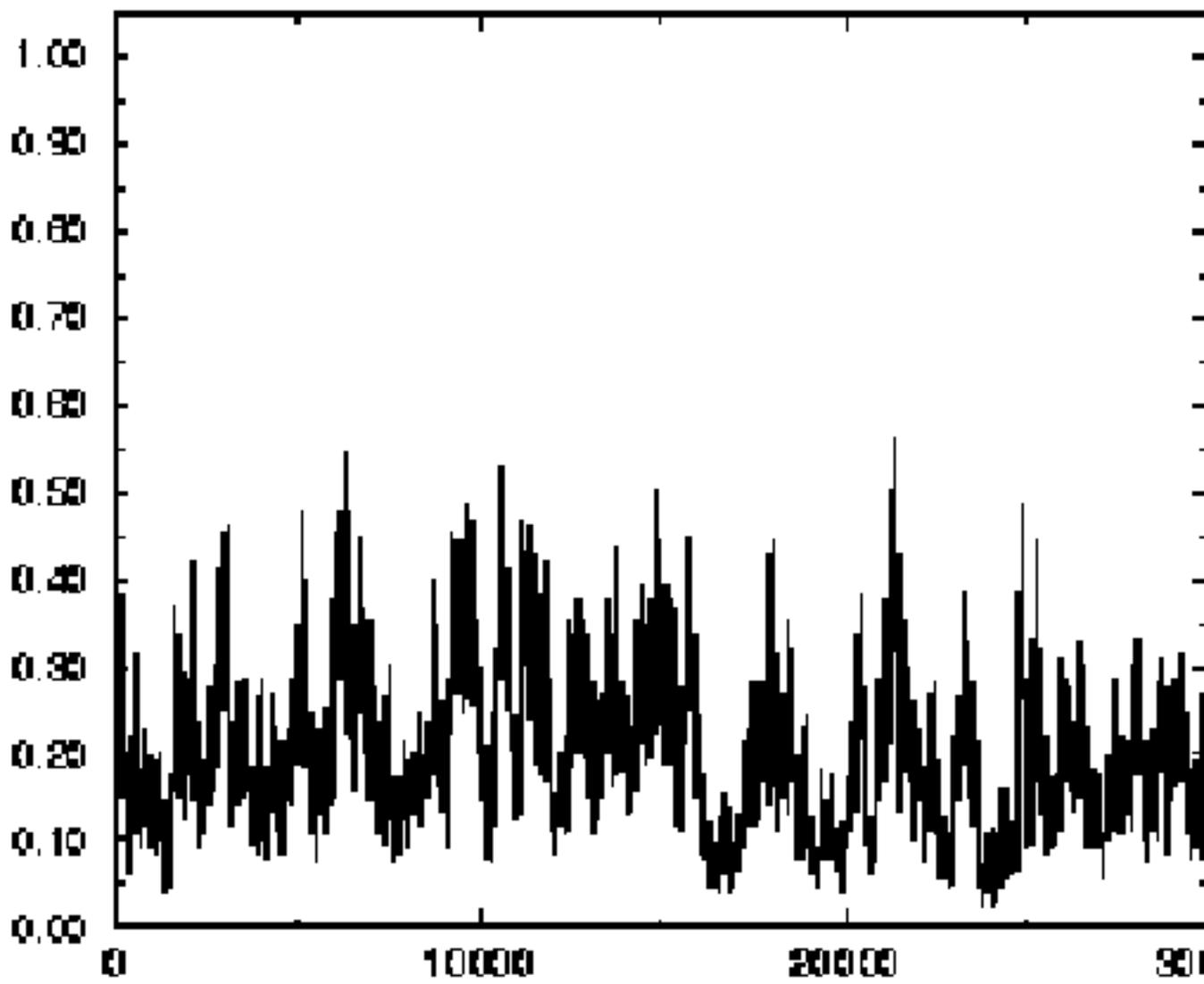
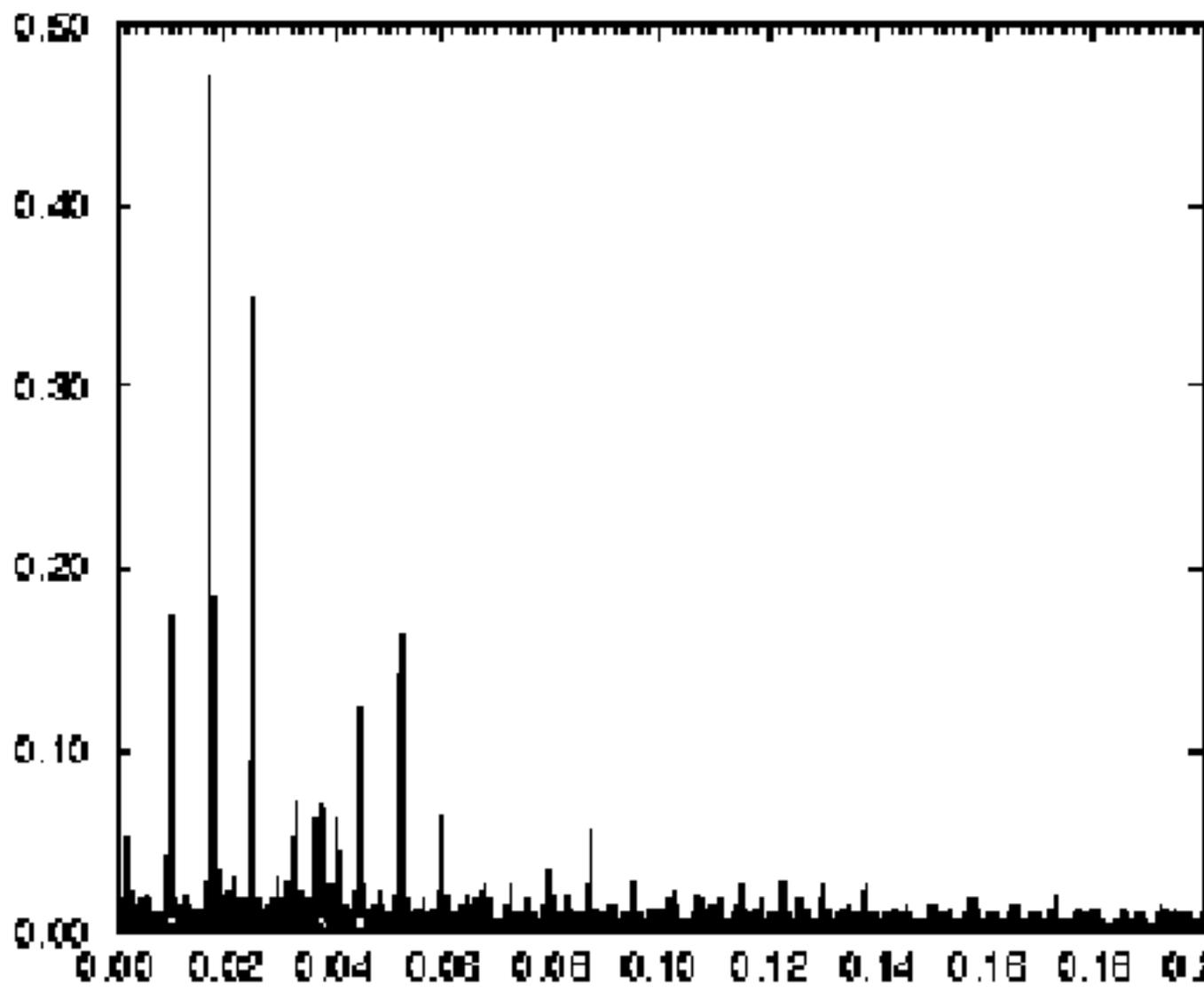
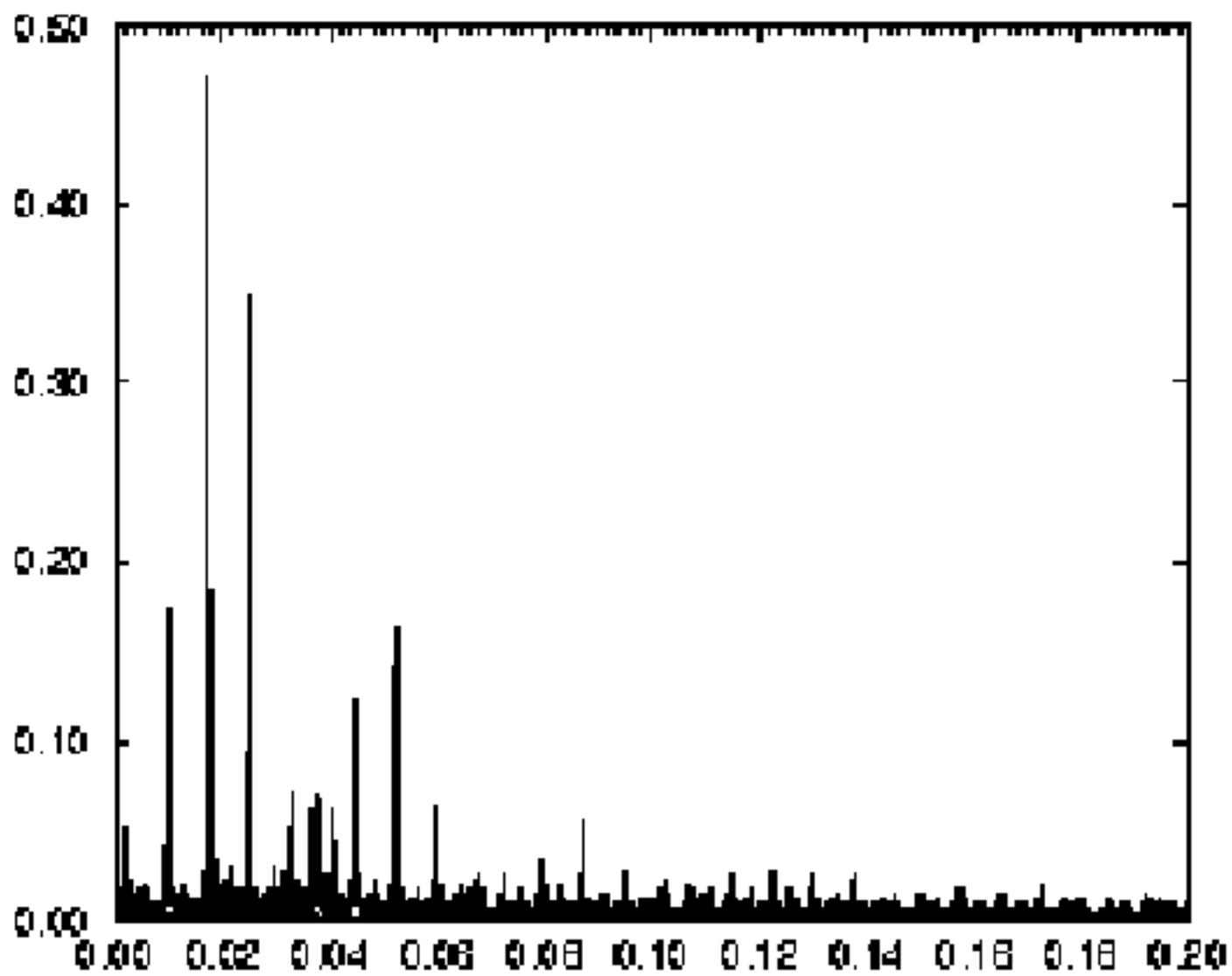
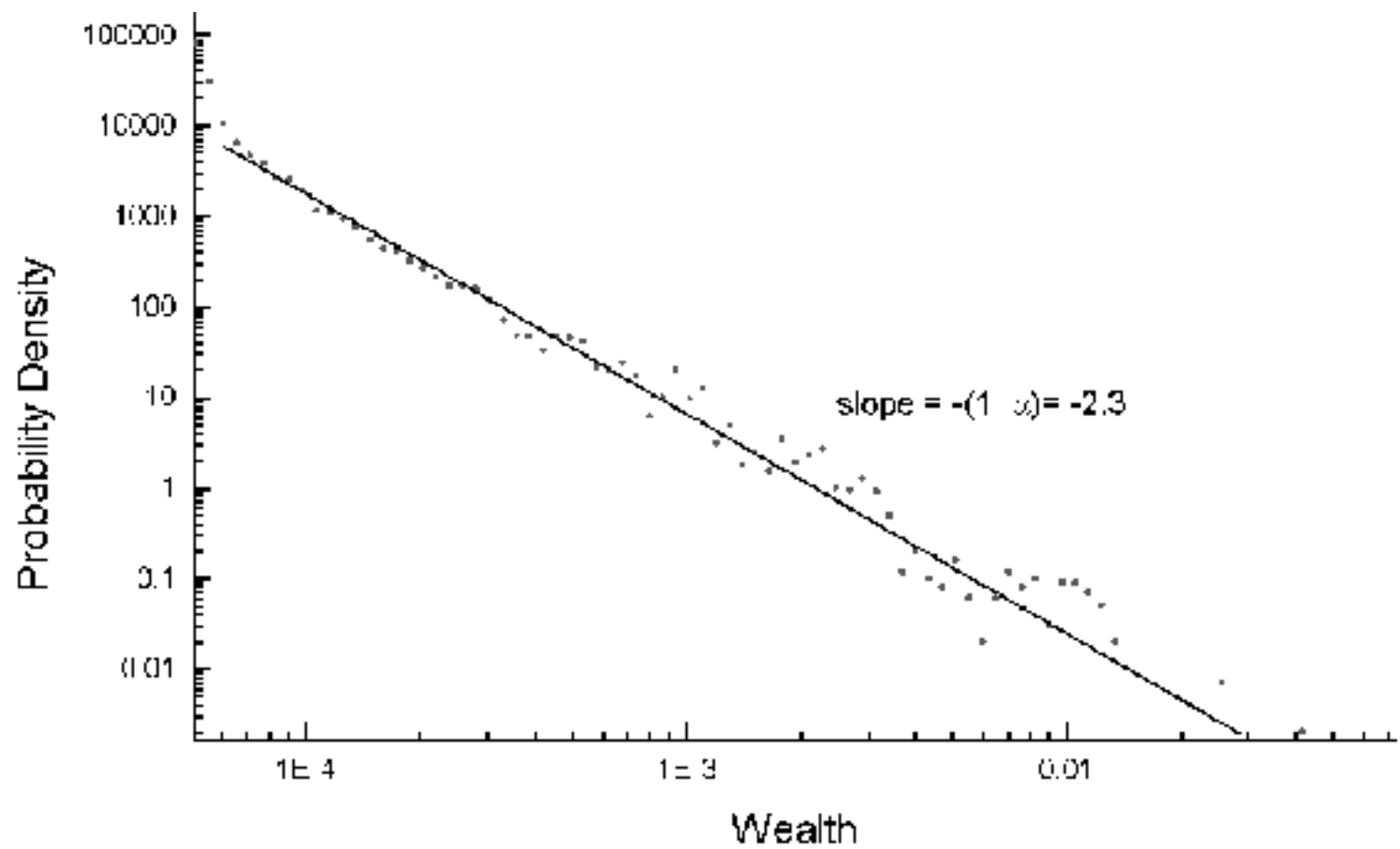
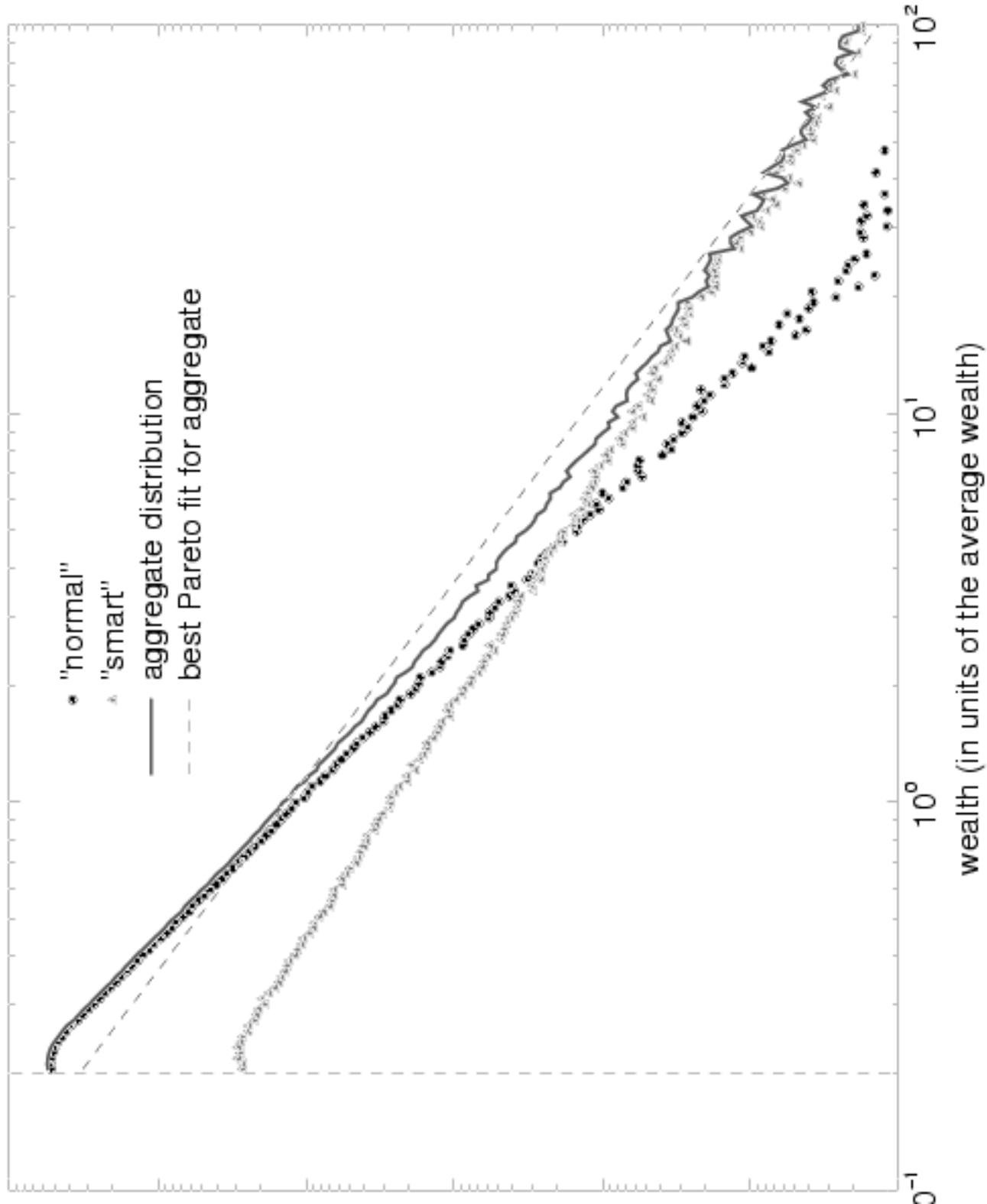
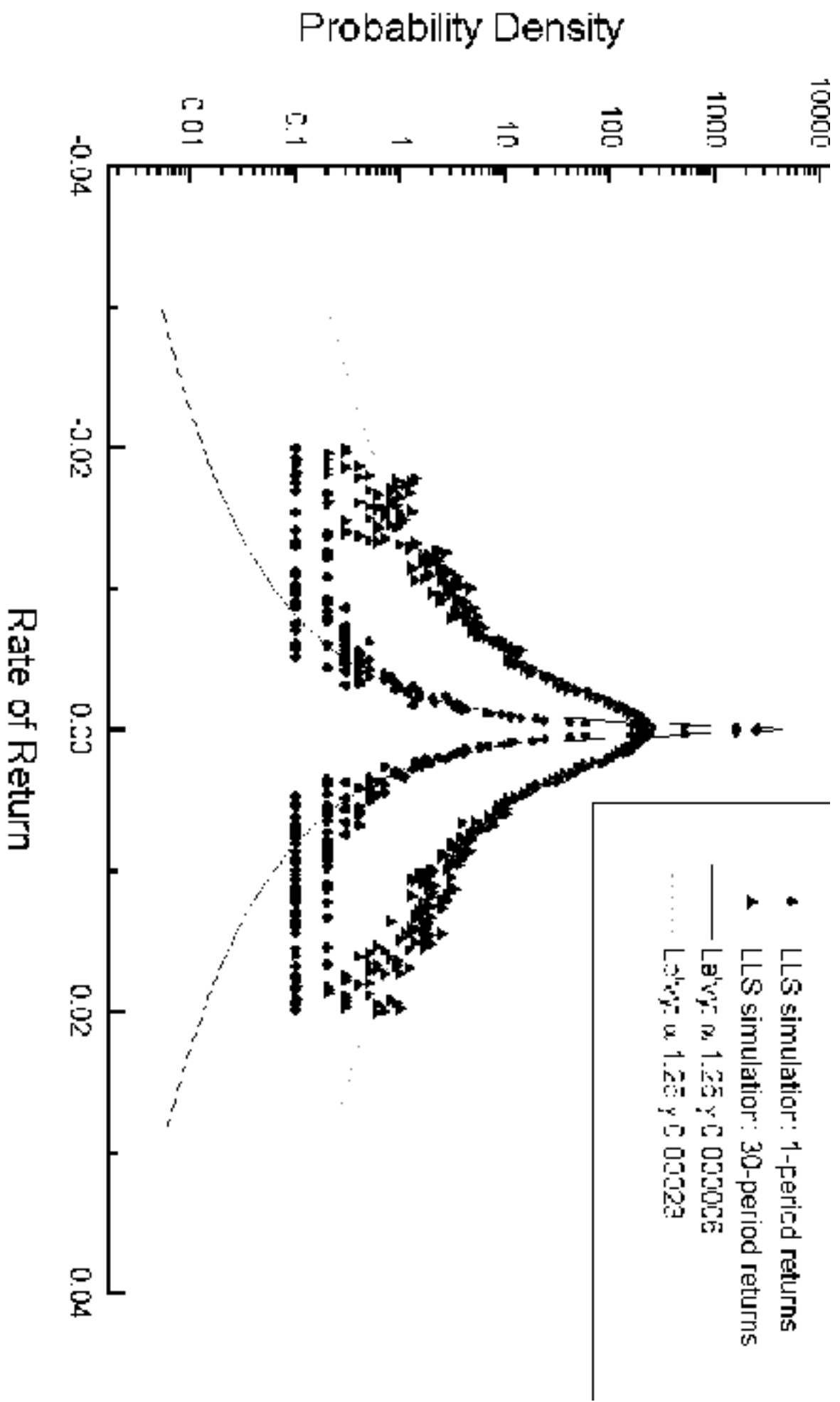
\end